# Polarization-Temperature Correlations in the Microwave Background [*]

ROBERT CRITTENDEN

*Joseph Henry Laboratory, Princeton University*
*Princeton NJ, 08544*

### Abstract

The polarization of the CMB offers a potentially valuable probe of the physics of the early universe, specifically of the velocity of the photon-baryon fluid on the last scattering surface. Here we discuss qualitatively the correlations that arise between the temperature anisotropies and polarization of the CMB and how they might be observed.

## 1 Introduction

Recent observations of large scale anisotropy in the cosmic microwave background [1] mark the beginning of a new epoch in experimental cosmology. These observations, when combined with rapidly improving measurements of the small scale anisotropy [2], should prove to be a powerful tool for distinguishing between competing cosmological models. It has recently become clear, however, that the CMB alone may not be sufficient to uniquely differentiate cosmological models [3] [4] [6] [5] and that other cosmological probes will be required to resolve this ambiguity.

One potentially powerful cosmological probe is the polarization of the microwave background [7] [8] [9]. Polarization arises from Thomson scattering off free electrons and is expected to be of order 10% or less of the temperature anisotropy [10]. Present experimental limits on the polarization are far from this level of sensitivity [11], but it is hoped that improvements in technology in the next generation of experiments may make such a detection possible. This makes it crucial that we understand the theoretical predictions for polarization.

---

[*]Talk presented at the 1994 Case-Western Reserve University Workshop "CMB Anisotropies Two Years After COBE," April 22-24.

Previous discussions of the polarization have focused on determining its auto-correlation function alone [10]. In these *Proceedings* we discuss how the temperature anisotropy and polarization are correlated in adiabatic models [12]. This correlation is inevitable because the polarization is created on the surface of last scattering only if there exists anisotropy in the incident photon temperature. A significant portion of the expected polarization signal is correlated with the temperature anisotropy, which makes it possible to construct a map of the correlated part of the polarization given a temperature anisotropy map. In addition, the temperature-polarization correlation may provide the clearest window for the detection of the polarization in the CMB.

## 2  Polarization

How is polarization in the CMB produced? When initially unpolarized light Thomson scatters from an electron, it becomes partially polarized. For example, light scattering off an electron at a 90° angle will be 100% linearly polarized perpendicular to the scattering plane. If the light incident on a scatterer is isotropic, then by symmetry, the scattered light will also be isotropic. However, if the incident light is anisotropic (specifically, if it has a quadrupole moment,) then the scattered light will become polarized.

Prior to recombination, the baryons and photons are tightly coupled and can be treated effectively as a single fluid. In this limit, the photon distribution is completely described by the density and velocity of this fluid. The photons and baryons do not remain perfectly coupled, however, and as recombination approaches, the photons begin to diffuse away from the baryons. Their mean free path is given by $ct_c = c/(\sigma_T n_e)$, where $\sigma_T$ is the Thomson cross section and $n_e$ is the density of free electrons. As the free electron density drops, $t_c$ grows and the fluid approximation breaks down.

In the fluid limit, the quadrupole of the incident photon phase space density is vanishingly small. As decoupling occurs, however, photons are able to free stream from surrounding regions. The phase space density in a region develops a quadrupole proportional to the divergence of the local velocity field, of order $t_c \dot{\delta}_\gamma \sim -t_c \nabla v_\gamma$, where $\delta_\gamma$ and $v_\gamma$ are the photon density and velocity perturbations respectively. This quadrupole in the incident light causes the scattered light to be polarized.

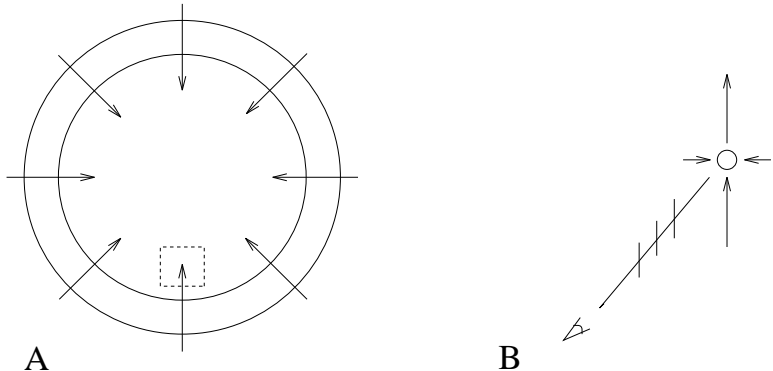

Figure 1: A growing potential well on the last scattering surface (1A.) The convergent velocity field induces a local quadrupole at points on the edge of the well. This quadrupole causes the light to be radially polarized about the well. (The boxed region is expanded in Fig. 1B.)

Consider a potential well (appearing as a cold spot on the microwave sky) which enters the horizon just prior to decoupling. Immediately after entering the horizon, gravitational forces cause the surrounding matter to fall into the well. (See figure 1A.) The converging velocity field induces a local quadrupole around points on the edge of the potential well, causing an excess of scattered light polarized radially about the cold spot (Figure 1B.) Thus the cold spot will be surrounded by a radial polarization pattern. Similarly, a potential hill (or hot spot) entering the horizon induces a diverging velocity field, which causes a deficit in the amount of radially polarized light about the hot spot, or a tangential polarization pattern.

For scales that were well within the horizon at last scattering, pressure forces come to dominate, causing the photon-baryon fluid to oscillate acoustically. The density and velocity both oscillate, but are out of phase with each other by 90°. An over-dense fluctuation can grow or decay, and so can be associated with either a converging or a diverging velocity field, depending on how long it has been in the horizon at last scattering. Thus, the sign of the cross correlation function oscillates as a function of the length scale of the perturbation. On these smaller scales, cold spots (and hot spots) can be associated with either tangential or radial polarization patterns.

The correlation function on a fixed angular scale receives contributions from perturbations with a range of physical length scales. This is because the relevant distance scale is the length of the perturbation projected onto the surface of last scattering rather than its actual length. That is, the angular scale that a given fluctuation contributes to depends on both its physical length and on the angle it makes with the last scattering surface. Because of this and the fact that the cross correlation function oscillates in sign as a function of wavelength, cancellations can occur between the fluctuations of different wavelengths that contribute to a given angular correlation. (Such cancellations do not occur in the calculation of auto-correlation functions.) As a result, not all of the polarization will be correlated with the temperature anisotropy. The greatest angular cross correlations occur for scales where the sign of the correlation does not change over the relevant window of physical scales. On scales where the velocity or density of the baryon-photon fluid does change sign, the cancellations will be greatest and there will be very little correlated polarization.

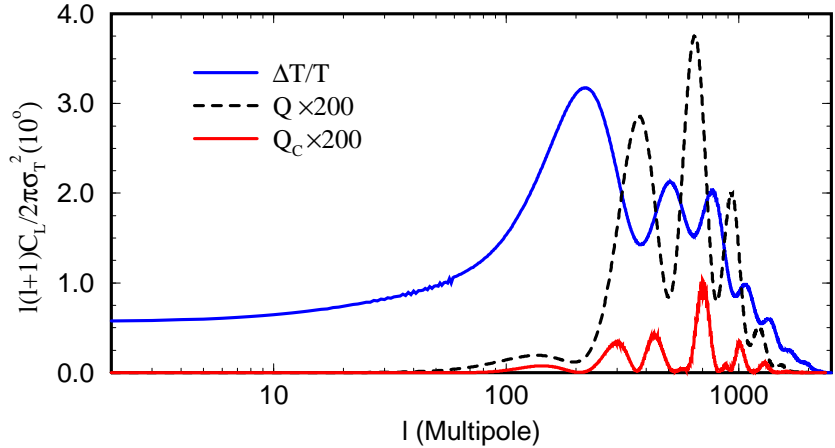

Figure 2: The power spectra of the temperature anisotropy, polarization and correlated polarization for a universe with a standard thermal history.

Figures 2 and 3 show the power spectra of anisotropies and polarization which result from scalar adiabatic fluctuations such as might arise from an inflationary epoch. We assume a flat universe with the baryon density constrained by nucleosynthesis ($\Omega_B h^2 = 0.0125$, $h = 0.5$) and the remaining mass to be cold dark matter. The calculations are performed by evolving the full Boltzmann equations from an early epoch, when there is no polarization, through recombination until the present. (We have used and extended the computational scheme of Bond and Efstathiou [10] [13] [14] [12].)

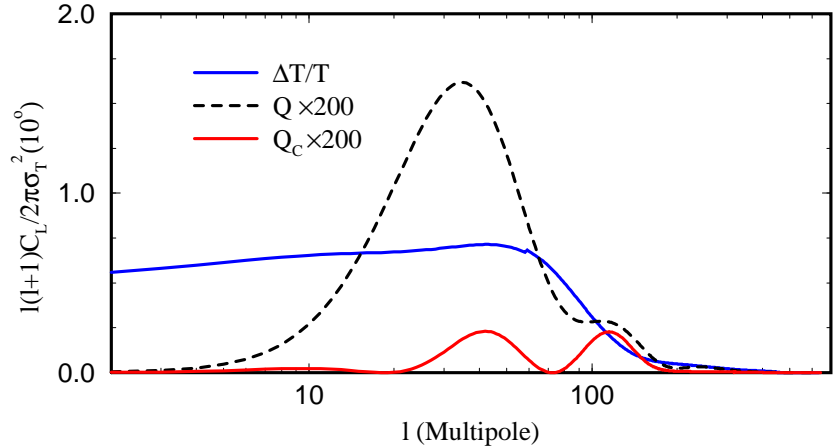

Figure 3: The power spectra of the temperature anisotropy, polarization and correlated polarization for a universe with no recombination.

Figure 2 shows the results for a universe with standard recombination at a red shift of approximately 1200. The polarization becomes substantial just after the "Doppler peak," on scales within the horizon at the time of recombination where scattering effects can begin to become important. (Remember that the term "Doppler peak" is a misnomer and that it doesn't actually represent the effect of the baryon velocities. In fact, the polarization is a much better probe of the baryon velocities on the last scattering surface.) Note that the peaks of the polarization power spectrum (which is scaled by a factor of 200 so that it can be seen on this plot) occur at the troughs of the anisotropy, resulting from the fact that the polarization and the anisotropy are out of phase. The lower curve is the correlated polarization power spectrum, which has peaks with twice the frequency of the total polarization. The area under the curves shows the variance of the total polarization is about seven times that of the correlated part.

Figure 3 shows similar plots for the opposite extreme thermal history, a universe which remains ionized until the present. All features are shifted to larger scales because last scattering occurs much later for these models. Anisotropies and polarization on very small scales are washed away by the thickness of the surface of last scattering. Note that the degree of polarization is quite similar to the standard recombination case, but that the coherence scale is much larger.

In a realistic experiment, one finds the degree of linear polarization by measuring the difference in amplitude of two orthogonal polarizations. (That is, $Q \equiv A_x - A_y$, where $A_i$ is the amplitude of the light polarized in the $i$ direction.) Because the polarization patterns will be radial or tangential, we expect that the temperature-polarization cross correlation function will have the form $\langle QT(\theta, \phi) \rangle = F(\theta) \cos 2\phi$. Here, $\theta$ is the length of the vector separating the polarization and temperature measurements (i.e. their angular separation) and $\phi$ is angle of this vector with respect to the polarization basis.

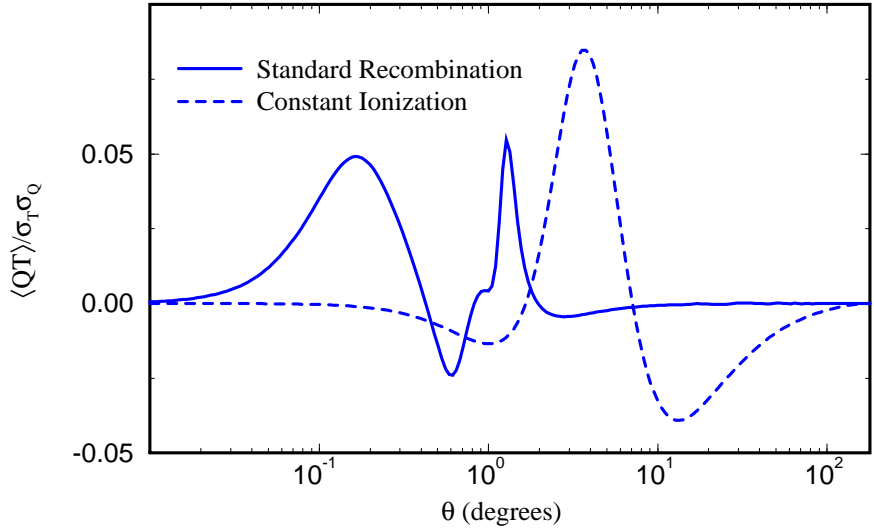

Figure 4: The temperature-polarization correlation function in units of $\sigma_T \sigma_Q$, where $\sigma_T^2$ and $\sigma_Q^2$ are the variances of the temperature anisotropy and polarization respectively.

Figure 4 shows the calculated $\theta$ dependence of the QT correlation function for both the standard and no recombination cases. They both vanish at small separations (as they must to prevent the correlation function from being multi-valued) and at large separations (because of causal constraints.) Both correlation functions oscillate between being positive and negative, though correlations for the no recombination case occur on much larger scales.

# 3  Experimental Implications

As we have mentioned, the polarization field can be expressed as the sum of two components, one correlated and one uncorrelated with the temperature field. The correlated component can be reconstructed from a temperature map, given some form for the cross correlation function. In the cases we have examined, the amplitude of the correlated polarization comprises only about a third of the total polarization signal. Since this is the case, in general a map of the correlated polarization is a poor predictor of the total polarization, except in a statistical sense. A peak of the total polarization map is more likely to occur at a peak of the correlated map than at a random point on the sky, but there will not be a one-to-one correspondence between between features of the two maps.

Although one expects the correlated polarization to be small, in principle it is easier to measure because temperature anisotropy measurements are expected to have a much greater signal to noise than polarization measurements. In addition, the temperature maps are likely to be more fully sampled than the polarization maps. The uncertainty in measurements of the auto-correlation function is proportional to the polarization detector noise squared, whereas the cross correlation uncertainty is linearly proportional to the polarization noise. If the noise in the polarization detector is much larger than the expected polarization signal, it becomes advantageous to search for correlated polarization, even though the signal is smaller. Because of this, the cross correlation function is more likely to be measured in a mapping experiment, whereas the polarization variance would be simpler to detect with an experiment which measures a few points with much higher accuracy.

In conclusion, the polarization-temperature correlations offer a new and potentially observable probe of the physics on the surface of last scattering. While these correlations are small, they are potentially within the grasp of the next generation of microwave measurements.

### Acknowledgements

I would like to thank David Coulson and Neil Turok, who collaborated on the work presented here [12]. I would also like to acknowledge many useful conversations with Ken Ganga, Jim Peebles, Paul Steinhardt and David Wilkinson. Finally, I thank Lawrence Krauss for his role in organizing this conference.